\documentclass[sigplan,screen,nonacm]{acmart}
\AtBeginDocument{%
  \providecommand\BibTeX{{%
    \normalfont B\kern-0.5em{\scshape i\kern-0.25em b}\kern-0.8em\TeX}}}
\usepackage{makecell}
\usepackage{listings}
\usepackage{color}
\usepackage{etoolbox}
\definecolor{codegreen}{rgb}{0,0.6,0}
\definecolor{codegray}{rgb}{0.5,0.5,0.5}
\definecolor{codepurple}{rgb}{0.58,0,0.82}
\definecolor{backcolour}{rgb}{0.95,0.95,0.92}
\definecolor{bluemunsell}{rgb}{0.0,0.53,0.70}
\definecolor{candypink}{rgb}{0.89,0.44,0.48}
\definecolor{britishracinggreen}{rgb}{0.0,0.26,0.15}
\definecolor{greencomments}{rgb}{0.13,0.54,0.13}
\lstdefinelanguage{WebAssembly}{
  sensitive=true,
  otherkeywords={},
  morekeywords=[1]{i32,f32,i64,f64},
  keywordstyle={[1]\color{violet}},
  morekeywords=[2]{0},
  keywordstyle={[2]\color{violet}},
  morekeywords=[3]{add,const},
  keywordstyle={[3]\color{bluemunsell}},
  morekeywords=[4]{},
  keywordstyle={[4]\color{candypink}},
  morekeywords=[5]{module, func, param, result, global, get_global, mut, set_global, export, import, memory, data, get_local, set_local, elem, table, call,call_indirect, type},
  keywordstyle={[5]\color{blue}},
  morekeywords=[6]{=,;},
  keywordstyle={[6]\color{britishracinggreen}},
  morekeywords=[7]{(,),[,],.},
  keywordstyle={[7]\color{black}},
  numberstyle=\tiny\color{black},
  rulecolor=\color{black},
  morecomment=[l][\itshape\color{greencomments}]{;;},
}
\lstdefinestyle{mystyle}{
    backgroundcolor=\color{backcolour},
    commentstyle=\color{codegreen},
    keywordstyle=\color{magenta},
    numberstyle=\tiny\color{codegray},
    stringstyle=\color{codepurple},
    basicstyle=\ttfamily\footnotesize,
    breakatwhitespace=false,
    breaklines=true,
    captionpos=b,
    keepspaces=true,
    numbers=left,
    numbersep=5pt,
    showspaces=false,
    showstringspaces=false,
    showtabs=false,
    tabsize=2
}
\usepackage{xspace}
\newcommand{\sysname}{ModTrans\xspace}
\begin{document}
\title{ModTrans: Translating Real-world Models for
Distributed Training Simulator}
\author{Yi Lyu}
\authornotemark[1]
\affiliation{%
  \institution{University of Wisconsin-Madison}
  \city{Madison}
  \state{WI}
  \country{USA}}
\email{ylyu76@wisc.edu}
\renewcommand{\shortauthors}{Lyu}
\newcommand{\codesm}[1]{\texttt{\small #1}}
\begin{abstract}
Large-scale distributed training has been a research hot spot in machine learning systems for industry and academia in recent years. However, conducting experiments without physical machines and corresponding resources is difficult. One solution is to leverage distributed training simulator. Current distributed training simulators like ASTRA-sim do not support the import from real-world developed models to the simulator, which poses challenges for ML researchers from using the simulator. Based on this challenge, We developed ModTrans, a translator supporting format translation from any real-world model to the ASTRA-sim simulator's input. It removes the barrier between machine learning experts and machine learning system researchers. The experiment results show that \sysname's cost is negligible.
\end{abstract}
\maketitle
\section{Introduction}
In recent years, machine learning training has become one of the most intensive workloads in data centers because of the increasing demands of machine learning applications and the complexity of the model itself. Giant models (or model experts) are considered the next-generation model, and training these models need intensive computational and communication resource. For example, a recent report from Google shows that the large model PaLM~\cite{palm} is trained with 81Tbps for each training step, and it takes about 6144 TPU to participate in the training process. 

Replicating such models' training on a large scale is extremely difficult because of the huge amount of network and computation usage. This happens in both industry and academia. For industry, it is costly to train models without enough testing, potentially increasing the cost of deployment and management. For academia, it is difficult for educational institutions to acquire a training environment that could replicate the experiments, which poses a barrier to researchers conducting distributed training research. 

Distributed training simulator is a solution to the problem above. ASTRA-sim is one of the main-stream machine learning simulators in this area. It simulates distributed training by defining the network topology and the computation model, letting developers replicate the distributed training environment in a single machine. It frees users from running the distributed training job in a real machine environment and occupies the resources. 

However, these simulators are far away from mature. It is due to the fact that distributed training behavior is complicated, and the computation and communication modeling are sophisticated. For example, ASTRA-sim only takes model description files as input, which must be written and configured manually. In the real scenario, the simulations are conducted by machine learning system researchers, and the ML models are taken from the ML developer, which are two groups of people. Therefore, we think it is not a reasonable design that needs to be decoupled. 

In this paper, we developed \sysname, a translator that would like any real-world models and extracted the necessary information needed from ASTRA-sim, making the ASTRA-sim supports simulation input directly taken from the model without manually configured. Our translator can also get classic models from the model zoo if users would like to use the simulator to simulate the model training, for example, ResNet50 and VGG16. Note that our translator is applicable to any simulator that takes layer-wise information as input to do simulation. 

\sysname removes the barrier between machine learning experts and machine learning system researchers. Our experiment results show that \sysname takes no more than 1 second to translate, and the translation results are identical to the ResNet50 model provided in the ASTRA-sim repository.

\section{Background}
In this section, we will discuss the background. The first subsection introduces the basic concepts like parallelism types in distributed deep learning training \cite{Efflex, vetrass, catp}. The second subsection introduces the architecture of state-of-the-art distributed training simulators and the interface layer facing the user. In the final subsection, we introduce the Open Neural Network Exchange (ONNX), an open-source framework for representing AI models.  
\subsection{Distributed Deep Learning Training}

Training in giant models introduces new challenges to data centers. How to train models effectively with hundreds of or thousands of computational devices (GPU/TPU) becomes an interesting question. Nvidia's state-of-the-art Megatron-LM~\cite{megatron-lm} model takes 3072 GPUs across 384 machines to train their large model with the combination of different training approach parallelism\cite{monom}. 

Data parallelism and Model parallelism are two main parallelism strategies that widely use in the data center because of the simplicity of the programmable model. Data parallelism is parallelization across multiple nodes by distributing data into different nodes, which can compute based on data in parallel. Model parallelism is another parallelization method that partitions the deep learning model\cite{safeguard} across multiple devices, within or across instances.   However, When models get larger and the architecture becomes more complex, we need to explore better solutions to accelerate the training efficiency. One approach is to combine different types of parallelism strategies within a model according to its characteristics. For example, some layers are too huge to fit into the rare GPU memory, and we need to split them into several partitions to train (model parallelism). 

Model parallelism introduces dependency problems among computational devices, potentially reducing the utilization of the computational devices. Gpipe~\cite{huang2019gpipe} and Pipedream~\cite{narayanan2019pipedream} introduce pipeline parallelism to increase device utilization by introducing the pipelining concept. For large model training, pipeline parallelism shows great performance improvement for huge data centers. The main insight from this line of work is trying to utilize workers' GPUs as much as possible, and they use different pipeline techniques to reduce the stall/bubble under naive execution.

At the same time, it is particularly difficult to do large-scale research academically on machine learning systems based on limited resources without industrial support. For example, BytePS~\cite{byteps} system uses 256 GPUs to prove its scalability, and getting the testbed with this amount of GPUs is already a huge challenge for academics. Moreover, system researchers are also exploring different deep-learning platform designs to accelerate the distributed training process. To find the best spot in the large design space, they usually need to try multiple different configurations and test the training performance. However, it is impossible to achieve that when running training jobs on the real testbed, since each training process takes such long GPU-hours. It motivates the need for distributed training simulator that can estimate the training performance to help researchers explore different design choices.

\subsection{ASTRA-sim}

\begin{figure}
    \centering
    \includegraphics[width=1.0\linewidth]{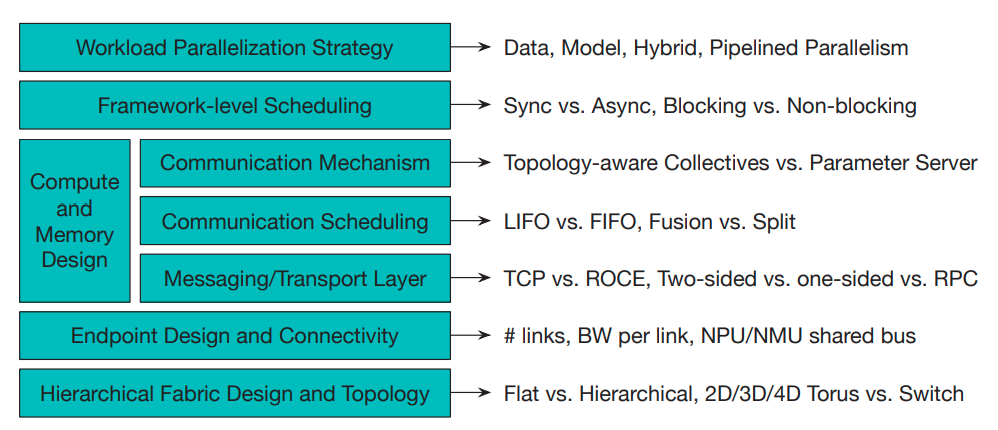}
    \caption{Deep Learning Platform Software (SW) and Hardware (HW) Design Space \cite{astrasim}}
    \label{fig:design-space}
    \vspace{6pt}
\end{figure}

As shown in Figure \ref{fig:design-space}, deep learning platforms have super complex hardware (HW) and software (SW) design space. On the hardware side, multiple accelerators are interconnected in different topologies. For accelerators, FPGA, GPU, TPUs, and other special ASICs are now widely used for deep learning training. For interconnection topology, there are different inter-package and intra-package fabrics with different bandwidths. In the software design space, there are different strategies for scheduling communication and computation. For communication, the protocol and topology-aware collectives can be configured by users. The communication scheduling can be configured as LIFO or FIFO. And the framework scheduling can be configured as synchronous or asynchronous, blocking or nonblocking. With increasingly large deep learning models, both academia and industry are trying to find the best distributed training platform design point inside this complex design space. \texttt{ASTRA-sim} \cite{astrasim} is a distributed training simulator designed to quickly simulate the training performance when navigating the design space.

\begin{figure}
  \centering
  \includegraphics[width=1.0\linewidth]{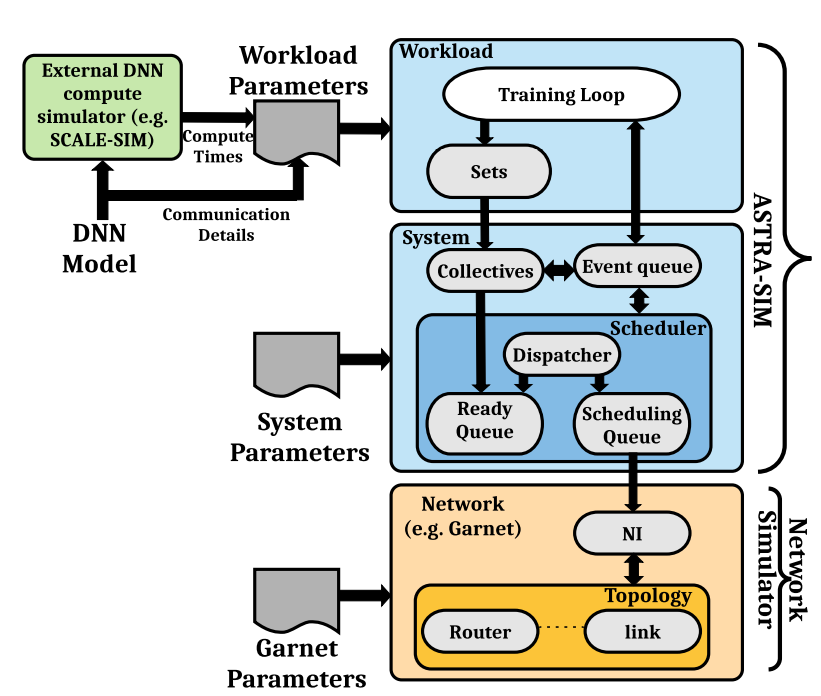}
  \captionof{figure}{Overview of ASTRAM-sim\cite{astrasim}}
  \label{fig:astra-sim}
\end{figure}

The \texttt{ASTRA-sim} is a simulator consisting of multiple layers (network layer, system layer, and workload Layer). Figure \ref{fig:astra-sim} shows the high-level overview of the simulator components. Its network layer uses Garnet \cite{agarwal2009garnet} to simulate the network behavior under different topologies and connectivity. It is mainly used to mainly simulate the communication stage when different nodes want to exchange information in the training process. \texttt{ASTRA-sim} is also exploring supporting another more common network simulator \textit{ns-3}, which is more flexible to configure different network protocols and communication patterns. 
The system layer provides topology-aware collective operations and generates traffic to the network layer. Also, it contains the scheduler component that pipelines the execution of the collectives across different links. Notice that the system layer deals with the logical topology while the network layer handles the actual network topology. The advantage here is that there could be a more flexible mapping between the logical and actual topologies. The workload layer runs the training loop algorithms for the specified deep learning models and generates the sets of data to be communicated during each iteration of training. There are two parts of workload parameters: \textit{compute time} and \textit{communication sizes}. For the compute time, it can be extracted from the DNN simulator and GPU simulator for each layer operation. For the communication size, it depends on the parallelism types and also the model itself.

\begin{figure}
  \centering
  \includegraphics[width=1.0\linewidth]{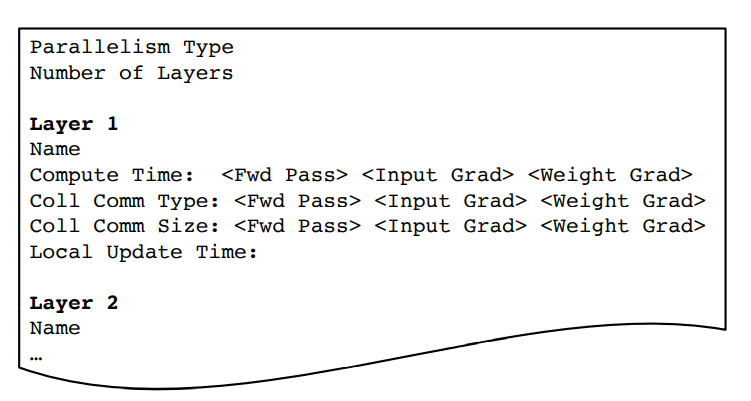}
  \captionof{figure}{DNN description file \cite{astrasim}}
  \label{fig:dnn-desc}
\end{figure}

The \textit{ASTRA-sim} simulator takes the DNN description file as input and generates the sets of data to be communicated at different training steps. As shown in Figure \ref{fig:dnn-desc}, the description file specifics the parallelism type and layer-wise information (computation \& communication). For each layer inside the deep learning model, the description file contains the computation time, communication type, communication size, and local update time. Since deep learning models have more and more layers, it's time-consuming for users to provide the DNN description file. Also, there could be potential errors when users calculate this information manually. It needs an automatic translator to generate this kind of DNN description file based on model specifications.

\begin{figure}
  \centering
  \includegraphics[width=1.0\linewidth]{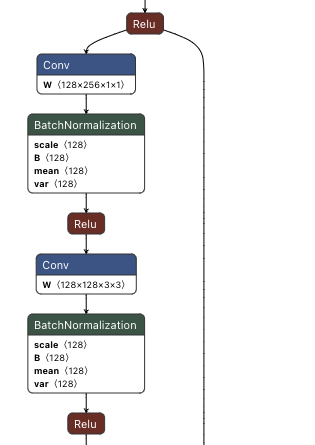}
  \captionof{figure}{Visualization on ONNX}
  \label{fig:vis-onnx}
\end{figure}

\subsection{ONNX}

Open Neural Network Exchange (ONNX) \cite{wiki:onnx} is an open-source format for AI models, both deep learning and traditional machine learning. The ONNX defines an extensible graph model and definitions of built-in operators and standard data types. It is widely supported by many frameworks, tools, and hardware. There are many industry collaborators that use the ONNX format and put it into production.

The core insight of ONNX is that deep learning with neural networks is usually expressed as computation over the dataflow graph. Some frameworks use static graphs (E.g., TensorFlow) while other frameworks use dynamic graphs (E.g., Pytorch) \cite{github:onnx}. Each framework provides its own interface to construct the dataflow graphs. ONNX is designed to provide a common intermediate representation (IR) to represent the dataflow graph. It can be compared to a programming language specialized in mathematical functions. All the necessary operators are already defined in the ONNX framework. There are five parts inside the model expressed in the ONNX language: input, output, node, initializer, and attributes. The input defines the type of data that should be fed into the model in ONNX. Similarly, the output defines the type of data predicted by the model in ONNX. The initializer is used to store the constant parameters in the model, which won't change based on the input. Nodes are the output of helper functions inside ONNX like \texttt{MatMul} and \texttt{Add}.   Listing \ref{lst:onnx} is a linear regression  that written in ONNX language \cite{github:onnx}
\begin{lstlisting}[language=python, label={lst:onnx}, caption={Linear Regression written in ONXX}]
def onnx_linear_regressor(X):
    "ONNX code for a linear regression"
    return onnx.Add(onnx.MatMul(X, coefficients), bias)
\end{lstlisting}

\begin{figure}
    \centering
    \includegraphics[width=0.8\linewidth]{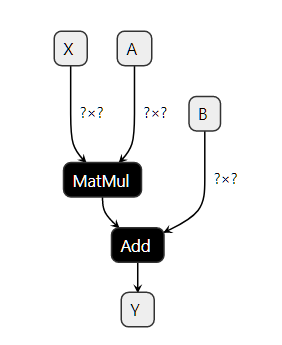}
    \caption{Linear Regression ONNX Graph}
    \label{fig:onnx-example-graph}
\end{figure}

As shown in Figure \ref{fig:onnx-example-graph}, the linear regression example in ONNX represents a graph, which transforms features into ONNX graph \cite{doc:onnx}. The \texttt{Add} and \texttt{MatMul} functions used in the Linear Regression model are corresponding nodes inside the ONXX graph. The arrow connecting different nodes shows the dataflow graph inside this deep learning model. 
The ONNX graph can be saved using \textit{protobuf} to serialize the graph into one single block. It is used to optimize the model size as much as possible. The generated format can be directly loaded into mainstream deep-learning models. Also, some visualization tools can take this binary format model as input to show the complete ONNX graph.

\section{Implementation Design}
In order to support real models in the ASTRA-sim simulator, we need to first understand the necessary information from the ONNX models. 
\subsection{Astra-SIM's Input}
The Astra-SIM simulator input includes three main sections, which are \textit{network}, \textit{system}, and \textit{workload}. The network and the system sections specify the network topology, communication patterns for training and scheduling policies, etc. \sysname focuses on the \textit{workload} section, which is the input file that describes the detail information of the input model, specifically, it takes layer communication size and layer computation time as input in this file.

The compute time of each layer is modeled by SCALE-sim~\cite{samajdar2018scale}, which is a low-level layer-wise computation time calculation simulator. However, the communication time, inferred from the model size, is manually extracted in the current ASTRA-sim design.

\subsection{\sysname Overview}
The design of \sysname is to remove the manual steps for extracting the model. In the current ASTRA-sim, if we want to simulate a model developed by ML experts, the simulator runner needs to know all the model information from the ML developer, which is not efficient and not well labor-divided. With \sysname, ML experts only need to develop their model in PyTorch, Tensorflow, or other Machine Learning frameworks that support ONNX. \sysname takes ONNX models and extracts the layer information for the ASTRA-sim. If developers want to use classic models, for example, ResNet50~\cite{resnet} and ~\cite{vgg}, \sysname also supports getting the models directly from the ONNX zoo~\cite{onnxzoo} by only giving the model name to \sysname. 

\subsection{\sysname Implementation}
\sysname is devleoped based on Python. The ONNX API is called to parse the ONNX. Since ONNX is in serialized binary format, we need to deserialize it before getting the data. Our evaluation results in Section~\ref{sec:eval} show that the deserialize cost is considerably small. After that, \sysname calculates the layer size based on the parsed data, for example, the number of parameters for each layer and data type.

\section{Evaluation}

\subsection{Experiment Setup}
Our experiments are all completed on a machine equipped with Intel(R) Xeon(R) CPU E5-2650 v3 @ 2.30GHz, and Python 3.8.10 installed. The ONNX models used in the following experiments come from the ONNX Model Zoo~\cite{onnxzoo}. We picked three classic machine learning models to demonstrate our results, which are ResNet50~\cite{resnet}, VGG16~\cite{vgg}, and VGG19~\cite{vgg}. In general, \sysname can be used in any machine with correct ONNX models.

\subsection{Execution Overhead}

\begin{figure}[!h]
  \centering
  \includegraphics[width=1\linewidth]{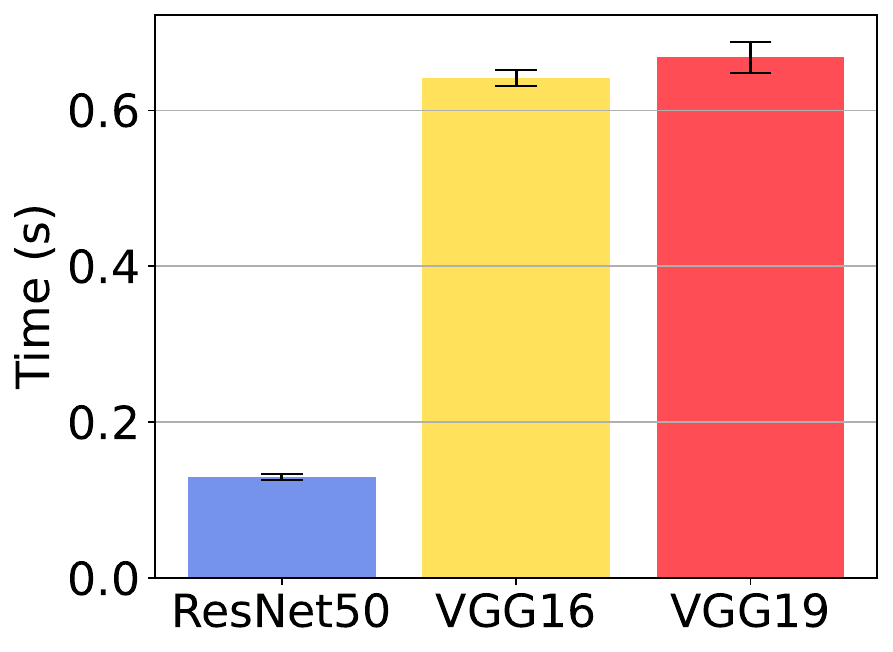}
  \captionof{figure}{Execution Time for \sysname}
  \label{fig:execu}
\end{figure}

The overhead of \sysname is negligible. Our experiments (Figure~\ref{fig:execu}) on three models show that the execution time for \sysname are nearly the same and all below a second. VGG16 and VGG19 take about 0.8 seconds, and ResNet50 takes only around 0.1 seconds. The variances are also small. 

Besides, since \sysname is not typically in the critical path, users can prepare the models offline. Therefore, we do not consider this is a cost that is needed to be concerned with.

\subsection{With different Models}

Our parser works for any model in the ONNX format.
In order to demonstrate the generality of our system, we run the \sysname and demonstrate the layer-by-layer information for VGG16 and VGG19 in Table~\ref{tab:vgg16} and Table~\ref{tab:vgg19}, respectively. This information includes but is not limited to \textit{Layer Name}, \textit{The Number of Variables}, \textit{Data Type}, and \textit{Model Size}. As shown in the tables, the layer-wise information is captured.

\subsection{Sanity Check}
To validate the results of our program, we compare the results with the example ResNet50 provided by the model ASTRA-SIm. Table~\ref{tab:resnet} shows the layer size of \sysname extracted ResNet50 model and ASTRA-SIm provided ResNet50 model. The results show that all the size information in each layer is identical. This demonstrates the correctness of \sysname and as a sanity check.

\begin{center}
\begin{table}[!h]
\begin{tabular}{|c|c|c|c|} 
 \hline
 Layer Name & Variables & Data Type & Model Size \\ [0.5ex] 
 \hline\hline
vgg16-conv0-weight & 1728 & FLOAT & 6912 \\
\hline
vgg16-conv1-weight & 36864 & FLOAT & 147456 \\
\hline
vgg16-conv2-weight & 73728 & FLOAT & 294912 \\
\hline
vgg16-conv3-weight & 147456 & FLOAT & 589824 \\
\hline
vgg16-conv4-weight & 294912 & FLOAT & 1179648 \\
\hline
vgg16-conv5-weight & 589824 & FLOAT & 2359296 \\
\hline
vgg16-conv6-weight & 589824 & FLOAT & 2359296 \\
\hline
vgg16-conv7-weight & 1179648 & FLOAT & 4718592 \\
\hline
vgg16-conv8-weight & 2359296 & FLOAT & 9437184 \\
\hline
vgg16-conv9-weight & 2359296 & FLOAT & 9437184 \\
\hline
vgg16-conv10-weight & 2359296 & FLOAT & 9437184 \\
\hline
vgg16-conv11-weight & 2359296 & FLOAT & 9437184 \\
\hline
vgg16-conv12-weight & 2359296 & FLOAT & 9437184 \\
\hline
vgg16-dense0-weight & 102760448 & FLOAT & 411041792 \\
\hline
vgg16-dense1-weight & 16777216 & FLOAT & 67108864 \\
\hline
vgg16-dense2-weight & 4096000 & FLOAT & 16384000 \\
\hline
\end{tabular}
\caption{Layer-by-layer sizes extracted from VGG16 ONNX model}
\label{tab:vgg16} 
\end{table}
\end{center}

\begin{center}
\begin{table}[!h]
\begin{tabular}{|c|c|c|c|} 
 \hline
 Layer Name & Variables & Data Type & Model Size \\ [0.5ex] 
 \hline\hline
vgg19-conv0-weight & 1728 & FLOAT & 6912 \\
\hline
vgg19-conv1-weight & 36864 & FLOAT & 147456 \\
\hline
vgg19-conv2-weight & 73728 & FLOAT & 294912 \\
\hline
vgg19-conv3-weight & 147456 & FLOAT & 589824 \\
\hline
vgg19-conv4-weight & 294912 & FLOAT & 1179648 \\
\hline
vgg19-conv5-weight & 589824 & FLOAT & 2359296 \\
\hline
vgg19-conv6-weight & 589824 & FLOAT & 2359296 \\
\hline
vgg19-conv7-weight & 589824 & FLOAT & 2359296 \\
\hline
vgg19-conv8-weight & 1179648 & FLOAT & 4718592 \\
\hline
vgg19-conv9-weight & 2359296 & FLOAT & 9437184 \\
\hline
vgg19-conv10-weight & 2359296 & FLOAT & 9437184 \\
\hline
vgg19-conv11-weight & 2359296 & FLOAT & 9437184 \\
\hline
vgg19-conv12-weight & 2359296 & FLOAT & 9437184 \\
\hline
vgg19-conv13-weight & 2359296 & FLOAT & 9437184 \\
\hline
vgg19-conv14-weight & 2359296 & FLOAT & 9437184 \\
\hline
vgg19-conv15-weight & 2359296 & FLOAT & 9437184 \\
\hline
vgg19-dense0-weight & 102760448 & FLOAT & 411041792 \\
\hline
vgg19-dense1-weight & 16777216 & FLOAT & 67108864 \\
\hline
vgg19-dense2-weight & 4096000 & FLOAT & 16384000 \\
\hline
\end{tabular}
\caption{Layer-by-layer sizes extracted from VGG19 ONNX model}
\label{tab:vgg19} 
\end{table}
\end{center}

\begin{center}
\begin{table}
\begin{tabular}{|c|c|c|} 
 \hline
 Layer Name & Extracted Model & ASTRA-SIM Model\\ [0.5ex] 
 \hline\hline
\hline
resnet-conv0 & 37632 & 37632 \\
\hline
resnet-stage1-conv0 & 16384 & 16384 \\
\hline
resnet-stage1-conv1 & 147456 & 147456 \\
\hline
resnet-stage1-conv2 & 65536 & 65536 \\
\hline
resnet-stage1-conv3 & 65536 & 65536 \\
\hline
resnet-stage1-conv4 & 65536 & 65536 \\
\hline
resnet-stage1-conv5 & 147456 & 147456 \\
\hline
resnet-stage1-conv6 & 65536 & 65536 \\
\hline
resnet-stage1-conv7 & 65536 & 65536 \\
\hline
resnet-stage1-conv8 & 147456 & 147456 \\
\hline
resnet-stage1-conv9 & 65536 & 65536 \\
\hline
resnet-stage2-conv0 & 131072 & 131072 \\
\hline
resnet-stage2-conv1 & 589824 & 589824 \\
\hline
resnet-stage2-conv2 & 262144 & 262144 \\
\hline
resnet-stage2-conv3 & 524288 & 524288 \\
\hline
resnet-stage2-conv4 & 262144 & 262144 \\
\hline
resnet-stage2-conv5 & 589824 & 589824 \\
\hline
resnet-stage2-conv6 & 262144 & 262144 \\
\hline
resnet-stage2-conv7 & 262144 & 262144 \\
\hline
resnet-stage2-conv8 & 589824 & 589824 \\
\hline
resnet-stage2-conv9 & 262144 & 262144 \\
\hline
resnet-stage2-conv10 & 262144 & 262144 \\
\hline
resnet-stage2-conv11 & 589824 & 589824 \\
\hline
resnet-stage2-conv12 & 262144 & 262144 \\
\hline
resnet-stage3-conv0 & 524288 & 2097152 \\
\hline
resnet-stage3-conv1 & 2359296 & 524288 \\
\hline
resnet-stage3-conv2 & 1048576 & 2359296 \\
\hline
resnet-stage3-conv3 & 2097152 & 1048576 \\
\hline
resnet-stage3-conv4 & 1048576 & 1048576 \\
\hline
resnet-stage3-conv5 & 2359296 & 2359296 \\
\hline
resnet-stage3-conv6 & 1048576 & 1048576 \\
\hline
resnet-stage3-conv7 & 1048576 & 1048576 \\
\hline
resnet-stage3-conv8 & 2359296 & 2359296 \\
\hline
resnet-stage3-conv9 & 1048576 & 1048576 \\
\hline
resnet-stage3-conv10 & 1048576 & 1048576 \\
\hline
resnet-stage3-conv11 & 2359296 & 2359296 \\
\hline
resnet-stage3-conv12 & 1048576 & 1048576 \\
\hline
resnet-stage3-conv13 & 1048576 & 1048576 \\
\hline
resnet-stage3-conv14 & 2359296 & 2359296 \\
\hline
resnet-stage3-conv15 & 1048576 & 1048576 \\
\hline
resnet-stage3-conv16 & 1048576 & 1048576 \\
\hline
resnet-stage3-conv17 & 2359296 & 2359296 \\
\hline
resnet-stage3-conv18 & 1048576 & 1048576 \\
\hline
resnet-stage4-conv0 & 2097152 & 8388608 \\
\hline
resnet-stage4-conv1 & 9437184 & 2097152 \\
\hline
resnet-stage4-conv2 & 4194304 & 9437184 \\
\hline
resnet-stage4-conv3 & 8388608 & 4194304 \\
\hline
resnet-stage4-conv4 & 4194304 & 4194304 \\
\hline
resnet-stage4-conv5 & 9437184 & 9437184 \\
\hline
resnet-stage4-conv6 & 4194304 & 4194304 \\
\hline
resnet-stage4-conv7 & 4194304 & 4194304 \\
\hline
resnet-stage4-conv8 & 9437184 & 9437184 \\
\hline
resnet-stage4-conv9 & 4194304 & 4194304 \\
\hline
resnet-dense0 & 8192000 & 8192000 \\
\hline
\end{tabular}
\caption{Layer-by-layer sizes extracted from ResNet50 ONNX model}
\label{tab:resnet} 
\end{table}
\end{center}

\section{Conclusion}
In this paper, we develop ModTrans that can extract the necessary information needed by ASTRA-sim from real-world models. It makes ASTRA-sim can support simulation directly from the model without being manually configured. The ModTrans translator can also get classic models from the model zoo if users would like to use the simulator to simulate the model training.

Since distributed training simulator is increasingly important and necessary, we believe that there will be more and more researchers exploring this field. It means that our ModTrans in the future can be generalized to any simulators that take layer-wide information as input to do simulation. 

In the evaluation, we show that ModTrans can translate the given ONNX models in less than 1 second. And the translated result is verified to be the same as the ResNet50 model provided in the official ASTRA-sim repository. ModTrans removes the barrier for users to run different deep learning models on the simulator. 
\bibliographystyle{ACM-Reference-Format}
\bibliography{sample-base}
\appendix

\end{document}